\documentclass[
    ,final            
 ,numberedheadings 
  ]
  {aipproc}

\layoutstyle{8x11double}

\usepackage[tight]{units}

\begin{document}

\title{Status of the CRESST Dark Matter Search}


\author{J.~Schmaler}{address={Max-Planck-Institut f\"ur Physik, F\"ohringer Ring 6, D-80805 M\"unchen, Germany}}
\author{G.~Angloher}{address={Max-Planck-Institut f\"ur Physik, F\"ohringer Ring 6, D-80805 M\"unchen, Germany}}
\author{M.~Bauer}{address={Eberhard-Karls-Universit\"at T\"ubingen, D-72076 T\"ubingen, Germany}}
\author{I.~Bavykina}{address={Max-Planck-Institut f\"ur Physik, F\"ohringer Ring 6, D-80805 M\"unchen, Germany}}
\author{A.~Bento}{address={Max-Planck-Institut f\"ur Physik, F\"ohringer Ring 6, D-80805 M\"unchen, Germany},altaddress={on leave from: Departamento de Fisica, Universidade de
Coimbra, P3004 516 Coimbra, Portugal}}
\author{A.~Brown}{address={Department of Physics, University of Oxford, Oxford OX1 3RH, United Kingdom}}
\author{C.~Bucci}{address={INFN, Laboratori Nazionali del Gran Sasso, I-67010 Assergi, Italy}}
\author{C.~Ciemniak}{address={Physik-Department E15, Technische Universit\"at M\"unchen, D-85747 Garching, Germany}}
\author{C.~Coppi}{address={Physik-Department E15, Technische Universit\"at M\"unchen, D-85747 Garching, Germany}}
\author{G.~Deuter}{address={Eberhard-Karls-Universit\"at T\"ubingen, D-72076 T\"ubingen, Germany}}
\author{F.~von~Feilitzsch}{address={Physik-Department E15, Technische Universit\"at M\"unchen, D-85747 Garching, Germany}}
\author{D.~Hauff}{address={Max-Planck-Institut f\"ur Physik, F\"ohringer Ring 6, D-80805 M\"unchen, Germany}}
\author{S.~Henry}{address={Department of Physics, University of Oxford, Oxford OX1 3RH, United Kingdom}}
\author{P.~Huff}{address={Max-Planck-Institut f\"ur Physik, F\"ohringer Ring 6, D-80805 M\"unchen, Germany}}
\author{J.~Imber}{address={Department of Physics, University of Oxford, Oxford OX1 3RH, United Kingdom}}
\author{S.~Ingleby}{address={Department of Physics, University of Oxford, Oxford OX1 3RH, United Kingdom}}
\author{C.~Isaila}{address={Physik-Department E15, Technische Universit\"at M\"unchen, D-85747 Garching, Germany}}
\author{J.~Jochum}{address={Eberhard-Karls-Universit\"at T\"ubingen, D-72076 T\"ubingen, Germany}}
\author{M.~Kiefer}{address={Max-Planck-Institut f\"ur Physik, F\"ohringer Ring 6, D-80805 M\"unchen, Germany}}
\author{M.~Kimmerle}{address={Eberhard-Karls-Universit\"at T\"ubingen, D-72076 T\"ubingen, Germany}}
\author{H.~Kraus}{address={Department of Physics, University of Oxford, Oxford OX1 3RH, United Kingdom}}
\author{J.-C.~Lanfranchi}{address={Physik-Department E15, Technische Universit\"at M\"unchen, D-85747 Garching, Germany}}
\author{R.~F.~Lang}{address={Max-Planck-Institut f\"ur Physik, F\"ohringer Ring 6, D-80805 M\"unchen, Germany}}
\author{M.~Malek}{address={Department of Physics, University of Oxford, Oxford OX1 3RH, United Kingdom}}
\author{R.~McGowan}{address={Department of Physics, University of Oxford, Oxford OX1 3RH, United Kingdom}}
\author{V.~B.~Mikhailik}{address={Department of Physics, University of Oxford, Oxford OX1 3RH, United Kingdom}}
\author{E.~Pantic}{address={Max-Planck-Institut f\"ur Physik, F\"ohringer Ring 6, D-80805 M\"unchen, Germany}}
\author{F.~Petricca}{address={Max-Planck-Institut f\"ur Physik, F\"ohringer Ring 6, D-80805 M\"unchen, Germany}}
\author{S.~Pfister}{address={Physik-Department E15, Technische Universit\"at M\"unchen, D-85747 Garching, Germany}}
\author{W.~Potzel}{address={Physik-Department E15, Technische Universit\"at M\"unchen, D-85747 Garching, Germany}}
\author{F.~Pr\"obst}{address={Max-Planck-Institut f\"ur Physik, F\"ohringer Ring 6, D-80805 M\"unchen, Germany}}
\author{S.~Roth}{address={Physik-Department E15, Technische Universit\"at M\"unchen, D-85747 Garching, Germany}}
\author{K.~Rottler}{address={Eberhard-Karls-Universit\"at T\"ubingen, D-72076 T\"ubingen, Germany}}
\author{C.~Sailer}{address={Eberhard-Karls-Universit\"at T\"ubingen, D-72076 T\"ubingen, Germany}}
\author{K.~Sch\"affner}{address={Max-Planck-Institut f\"ur Physik, F\"ohringer Ring 6, D-80805 M\"unchen, Germany}}
\author{S.~Scholl}{address={Eberhard-Karls-Universit\"at T\"ubingen, D-72076 T\"ubingen, Germany}}
\author{W.~Seidel}{address={Max-Planck-Institut f\"ur Physik, F\"ohringer Ring 6, D-80805 M\"unchen, Germany}}
\author{L.~Stodolsky}{address={Max-Planck-Institut f\"ur Physik, F\"ohringer Ring 6, D-80805 M\"unchen, Germany}}
\author{A.~J.~B.~Tolhurst}{address={Department of Physics, University of Oxford, Oxford OX1 3RH, United Kingdom}}
\author{I.~Usherov}{address={Eberhard-Karls-Universit\"at T\"ubingen, D-72076 T\"ubingen, Germany}}
\author{W.~Westphal}{address={Physik-Department E15, Technische Universit\"at M\"unchen, D-85747 Garching, Germany}, altaddress={Deceased}}

\begin{abstract}
The CRESST experiment aims for a detection of dark matter in the form of WIMPs. These particles are expected to scatter elastically off the nuclei of a target material, thereby depositing energy on the recoiling nucleus. CRESST uses scintillating CaWO$_4$ crystals as such a target. The energy deposited by an interacting particle is primarily converted to phonons which are detected by transition edge sensors. In addition, a small fraction of the interaction energy is emitted from the crystals in the form of scintillation light which is measured in coincidence with the phonon signal  by a separate cryogenic light detector for each target crystal.  The ratio of light to phonon energy permits the discrimination between the nuclear recoils expected from WIMPs and events from radioactive backgrounds which primarily lead to electron recoils. CRESST has shown the success of this method in a commissioning run in 2007 and, since then, further investigated possibilities for an even better suppression of backgrounds. Here, we report on a new class of background events observed in the course of this work. The consequences of this observation are discussed and we present the current status of the experiment. 
\end{abstract}

\maketitle


\section{Introduction}

It remains one of the most pressing challenges of astroparticle physics to clarify the nature of dark matter by a direct detection of the corresponding particles. A theoretically well-motivated candidate for those particles are WIMPs (weakly interacting massive particles), and there is an ongoing effort  of many experiments to directly detect them via their elastic scattering off the nuclei of a target. In such an interaction, a tiny amount of energy (typically of the order of \unit[10]{keV}) is transferred to the recoiling nucleus which can be detected. Cryogenic detectors with their low threshold and excellent energy resolution are well suited for this task. An additional challenge arises from the very low event rates anticipated (less than 10 events per kilogram of target material and year of measuring time) due to the small WIMP-nucleus scattering cross section. This requires a very efficient suppression of background events. 

CRESST (cryogenic rare event search with superconducting thermometers)  is one of the experiments currently aiming for such a direct detection of WIMPs. It is located in the Gran Sasso underground laboratory in Italy. 

\section{The CRESST Experiment}


CRESST uses scintillating CaWO$_4$ crystals as target for WIMP scatterings. They have a cylindrical shape (\unit[4]{cm} diameter, \unit[4]{cm} height) and weigh about \unit[300]{g} each. They are operated as cryogenic calorimeters at temperatures of about \unit[10]{mK}. 

When a particle interaction takes place, the deposited energy is mostly converted into phonons which can be detected (hence we also refer to the crystals as \emph{phonon detectors}). To this end, each crystal has a thin tungsten film evaporated on it which is operated as a transition edge sensor (TES). The film temperature is stabilized within the transition from the normal to the superconducting state, where the electrical resistance of the film strongly depends on its temperature. Thus, when phonons are absorbed in the film and temporarily heat it up, the film resistance rises. This signal is read out by SQUID-based electronics, ultimately resulting in a voltage pulse for each particle interaction. The height of this pulse can be used as a measure for the energy deposited in the crystal. CRESST has shown that, with this detector technology, it is possible to reach very low energy thresholds of about \unit[1]{keV} and an excellent energy resolution of about \unit[300]{eV} (FWHM) at low energies (\citep{Angloher2009_run30}).

In addition to the creation of phonons, a small fraction of the interaction energy (order of 1\%) is emitted from the crystal in the form of scintillation light. To detect this light, each crystal is paired with a separate cryogenic light detector made from a sapphire wafer (\unit[4]{cm} diameter, \unit[0.4]{mm} thickness) with a \unit[1]{\textmu m}  silicon layer on one side, acting as photon absorber. Similar to the crystals, the light detectors have a thin evaporated tungsten film which is operated as a transition edge sensor to read out the light signal.

\begin{figure}
	\includegraphics[width=0.8\linewidth]{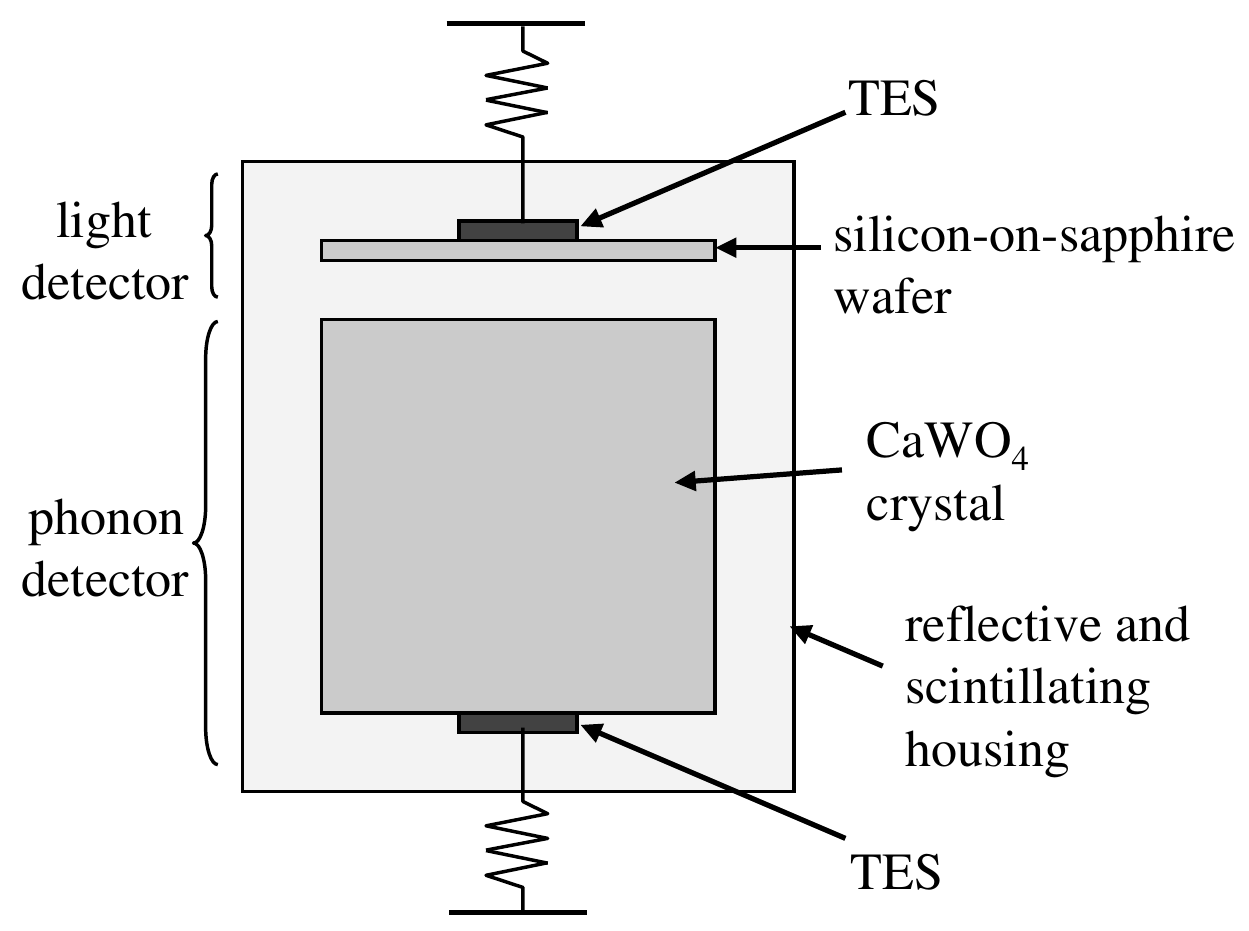}
	\caption{Schematic drawing of a CRESST detector module consisting of the target crystal and an independent light detector. Both are read out by transition edge sensors (TES) and enclosed in a common reflective and scintillating housing.}
	\label{fig:module}
\end{figure}

A crystal and the corresponding light detector form a so-called detector module as shown in Fig.\,\ref{fig:module}.
Both detectors of such a module are enclosed in a common reflective housing in order to collect as much scintillation light as possible. As a reflector, a polymeric foil was chosen which also shows scintillation properties. This is an important measure in oder to suppress WIMP-like background events due to surface contaminations with $\alpha$-emitters (\citep{Westphal2008_recoils}), as will be explained below.

For each particle interaction, a detector module yields two signals (from the phonon and the light detector). From these we define the \emph{light yield} of an event as the amount of energy in the light detector divided by the energy in the phonon detector and normalize it such that \unit[122]{keV}-gamma interactions have a light yield of 1. Electron recoils in general (caused by electrons or gammas interacting in the crystal) then show a light yield of approximately 1.  Compared to such events, $\alpha$-particles are found to give a factor of about 5 less light. Neutrons, on the other hand, mainly transfer measurable energy to the light oxygen nuclei in the CaWO$_4$ crystal (for kinematical reasons) and such oxygen recoils show a light yield of about 1/10. Finally, WIMPs are expected to mainly scatter off the heavy tungsten nuclei due to the properties of the coherent elastic scattering cross section, and these tungsten recoils were measured to have a light yield of about 1/40 (\citep{Bavykina2007_QF}). Thus the light yield is a powerful parameter for the discrimination between  potential WIMP events and the dominant radioactive backgrounds. 


The current CRESST setup can accommodate up to 33 detector modules which are mounted in a common support structure cooled to millikelvin temperatures by a dilution cryostat.
The detector volume is surrounded by low background copper and lead as inner shielding, which, in turn, is enclosed in a gas-tight box to avoid penetration of radon inside the shielding. In addition, a muon veto as well as a neutron shielding made from polyethylene 
have been installed to reduce the background interaction rate in the detectors as much as possible. A detailed description of the experiment can be found in  \citep{Angloher2009_run30}.


\section{Commissioning Run}

\begin{figure}
	\includegraphics[width=0.85\linewidth]{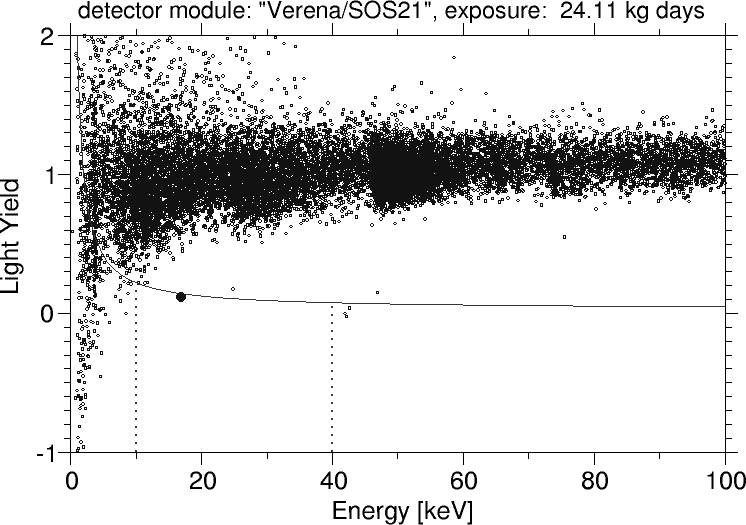}
	\hspace{1cm}
	\includegraphics[width=0.85\linewidth]{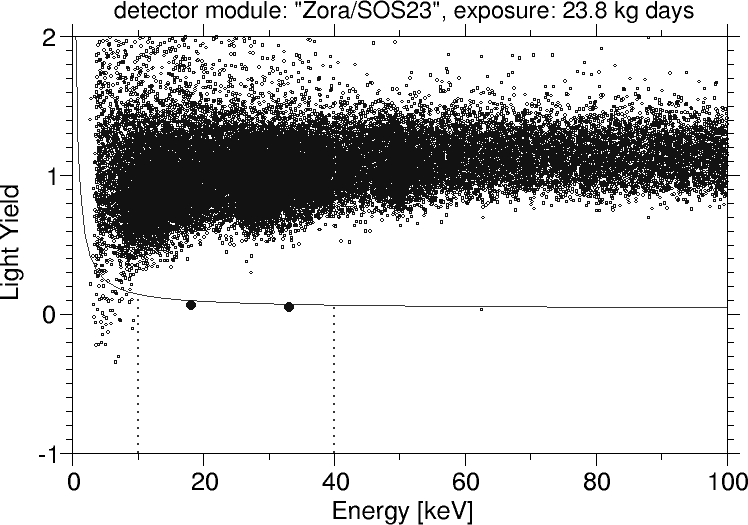}
	\caption{The background data of the two detector modules operated during the commissioning run in 2007. In total, three events were found in the acceptance region of tungsten recoils. }
	\label{fig:comm_results}
\end{figure}

In the year 2007, an extended commissioning run with the setup described above was carried out. Although this was mainly for optimization purposes, it was already possible to extract significant limits on dark matter interactions from the data collected by two detector modules with a total exposure of about \unit[48]{kg\,days}. Fig.\,\ref{fig:comm_results} shows these data in the light yield-energy-plane. The band of electron and gamma events centered around a light yield of 1 is clearly visible and below the solid lines we expect 90\% of the tungsten recoil events. The acceptance region which we consider for the dark matter analysis lies below these lines and extends from energies of \unit[10]{keV} (where the discrimination of nuclear recoils becomes efficient) to \unit[40]{keV} (roughly the maximum energy which a WIMP can transfer to a tungsten nucleus due to the nuclear form factor). 

A total of three candidate events was found in this "tungsten recoils" acceptance region, which, given the current dark matter limits, we consider as background events rather than WIMPs. From this, using standard assumptions for the dark matter distribution in our galaxy (cp.\ for example \citep{Donato1998_halorotation}), we derive an upper limit on the coherent WIMP-nucleon scattering cross section  which is as low as \unit[$4.8 \times 10^{-7}$]{pb} for an assumed WIMP mass of \unit[50]{GeV/c$^2$}. The details of this analysis are presented in~\citep{Angloher2009_run30}.

Further improvement of this limit requires an understanding and suppression of the remaining background events in the signal acceptance region. While the nature of these events could not be completely clarified, the two most likely causes seemed to be:
\begin{itemize}
	\item remaining neutrons that had traversed the shielding or had been emitted inside, e.g.\ by radioactivity intrinsic to other, non-operational crystals
	
	\item nuclear recoil events due to surface $\alpha$-decays in the surrounding of the crystals, where the $\alpha$-particle escapes detection and the recoiling heavy nucleus hits the crystal and mimics a WIMP event. 

	In order to veto this class of events, most of the detector surrounding is made from scintillating material as described above. When hit by the $\alpha$-particle, this material emits additional scintillation light in coincidence with the recoil in the crystal, thus increasing the total light yield of the event and moving it out of the WIMP acceptance region. In particular, the bronze clamps holding the crystals were covered with a reflecting and scintillating polymeric foil during the commissioning run for this reason, but some small areas could not be reached, leaving this mechanism a possible source of the WIMP candidate events.
\end{itemize}

In the light of these potential background sources, two main developments were conducted after the commissioning run:
\begin{enumerate}
	\item New phonon detectors with a higher light output were developed. This was realized by not directly evaporating the tungsten film onto the large crystal but on a separate smaller crystal which was then glued to the actual target crystal. This avoids a degradation of the light output of the target crystal during the evaporation process. The higher amount of scintillation light then allows for a considerably better discrimination of nuclear recoils and may, in particular, help to more reliably identify neutrons in the future. We report on these glued detectors and their performance in \citep{Kiefer2009_LTD}.
	
	\item In order to further suppress WIMP-like events due to surface $\alpha$-decays, new holding clamps for the crystals were manufactured which were completely covered with scintillating epoxy and had no non-scintillating surface areas anymore.  
\end{enumerate}

\section{2008 data taking and Status}

In the subsequent run between August and December 2008, 9 detector modules could be reliably operated, one of them with a glued target crystal as described above.
Several detector modules were equipped with the new, completely scintillating clamps, while, for comparison, other modules were again mounted using the foil-covered clamps of the commissioning run. Finally, the clamps of one crystal were left completely without any scintillating coverage in order to estimate the effect of this on the number of background events. 

\begin{figure}
	\includegraphics[width=0.85\linewidth]{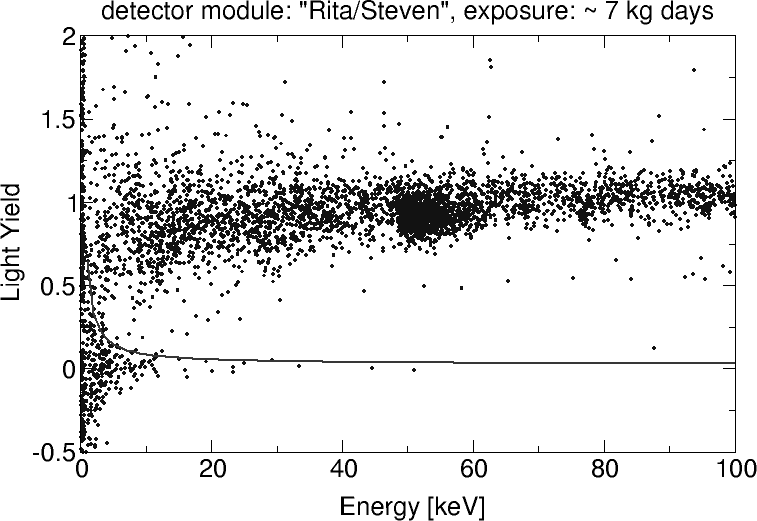}
	\hspace{1cm}
	\includegraphics[width=0.85\linewidth]{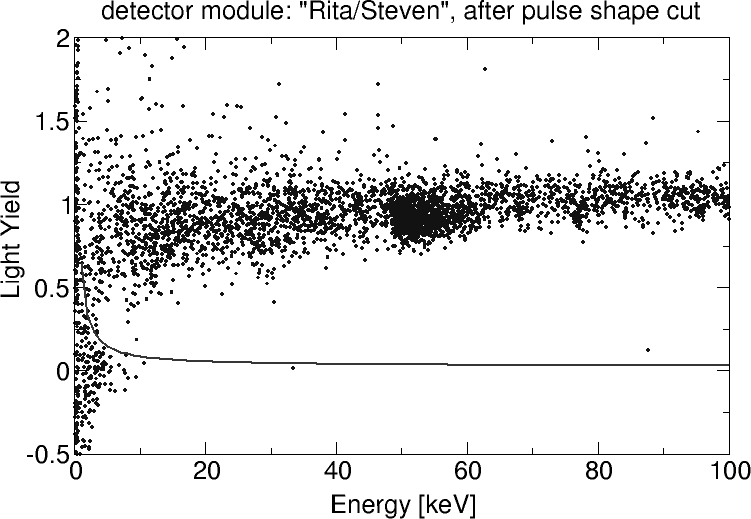}
	\caption{Data recorded with one detector module during the 2008 run. After a standard analysis, a considerable number of no-light events was found in the WIMP acceptance region  (below the solid line, where 90\% of the tungsten recoil events are expected) and in the relevant energy interval between 10 and 40~keV (left). Part of these events has a different pulse shape from normal particle pulses and can thus be rejected. In this example, this leaves the acceptance region almost clean (right).}
	\label{fig:rita_dark}
\end{figure}

Searching for events in the WIMP acceptance region, it turned out that almost all detector modules had seen events where no scintillation light could be detected, with the rate of such events varying considerably between the different detectors. Fig.\,\ref{fig:rita_dark} shows the most prominent example with several ten such events detected in about \unit[7]{kg\,days} of exposure. For the other modules, typically less than 10 such events were found in a similar exposure. The crystal of the shown module was held by the new epoxy-covered clamps.  

Such statistics of no-light events allowed a closer inspection of their properties. In particular it turned out that part of the no-light events had a slightly different pulse shape (longer decay time) from normal particle pulses. These can therefore be rejected by simple pulse shape cuts (cp.\ Fig.\ \ref{fig:rita_dark}, right). Nevertheless, there exist other no-light events with particle-like shape which cannot be cut this way. Generally, the rejection efficiency of the shape cut for no-light events depends very much on the individual detector.  

Looking at the time distribution of the no-light events, a significant decrease of their rate with time was observed. 

Remarkably, the number of no-light events was particularly low (only one) for the detector module equipped with uncovered metal holding clamps, while the other detectors (with foil- or epoxy-covered clamps) generally showed higher rates.

These facts suggest a detector effect as origin of the no-light events rather than ordinary particle interactions. We believe that the most likely source are stress-relaxation events which may occur at the contact area between the crystals and their holding clamps due to the rather tight clamping. Such events can happen in two different ways:
\begin{itemize}
 \item The relaxation can take place in the crystal itself in the form of \emph{micro-cracks}. Earlier experience has shown that such an energy release leads to pulses with a particle-like shape (\citep{Astrom2006_cracks}).  
 \item It may, however, also occur in the holding clamps, in particular in the scintillating plastic layer covering the surface. In this case, one would expect pulses with different shape due to the slow propagation of the created phonons in plastic.   
\end{itemize}
We assume that, in our case, both types of relaxation events exist, with the plastic-covered clamps being more vulnerable to the second type than the pure metal ones.

Given this explanation, two major changes of the holding clamps were performed after the 2008 run:
\begin{itemize}
	\item New clamps were manufactured from thinner and less stiff bronze material to reduce the pressure they exert on the crystals and thus to avoid micro-cracks.  
	\item The new clamps were only covered with a silver layer to make them reflecting, but any kind of scintillating plastic coverage was avoided. This requires even more care not to contaminate the surfaces with $\alpha$-emitters in order to avoid the dangerous nuclear recoil background. 
\end{itemize}


With all crystals equipped with the modified clamps, CRESST has started a new run in June 2009. Currently, 9 detector modules are fully operational, including one crystal made from the alternative target material ZnWO$_4$. Some more detectors are expected to come into operation as the cryostat is cooling further. By the time of writing, the situation with no-light events could not yet be clarified. We will report on this at a later stage.



\begin{theacknowledgments}
This work was partially supported by funds of the DFG (Transregio 27, "Neutrinos and Beyond"), the Munich Cluster of Excellence ("Origin and Structure of the Universe") and the Maier-Leibnitz-Laboratorium (Garching).
\end{theacknowledgments}



\bibliographystyle{aipproc}   

\bibliography{bibliography}

\IfFileExists{\jobname.bbl}{}
 {\typeout{}
  \typeout{******************************************}
  \typeout{** Please run "bibtex \jobname" to optain}
  \typeout{** the bibliography and then re-run LaTeX}
  \typeout{** twice to fix the references!}
  \typeout{******************************************}
  \typeout{}
 }

\end{document}